%% file: main.tex
\documentclass[12pt]{article}
\usepackage[a4paper, total={6in, 8in}]{geometry}
\usepackage{graphicx} 
\usepackage{amsmath,bm}
\usepackage{amssymb}
\usepackage[
colorlinks=true,
urlcolor=blue,
linkcolor=blue,
citecolor=blue,
pdfencoding=auto, psdextra
]{hyperref}
\usepackage[export]{adjustbox}
\usepackage{cancel}
\usepackage{array}
\usepackage{centernot}

\usepackage{authblk}
\usepackage{soul}

\usepackage[dvipsnames]{xcolor}
\definecolor{RubineRed}{HTML}{ED017D}
\definecolor{ao(english)}{rgb}{0.01, 0.75, 0.24}
\definecolor{ballblue}{rgb}{0.13, 0.67, 0.8}
\definecolor{amethyst}{rgb}{0.6, 0.4, 0.8}
\definecolor{cinnabar}{rgb}{0.89, 0.26, 0.2}

\usepackage{setspace}

\newtheorem{definition}{Definition}

\title{Unified Causality Analysis Based on the Degrees of Freedom}

\author[a,c]{Andr\'as Telcs}
\author[a,b]{Marcell T. Kurbucz}
\author[a,b]{Antal Jakov\'ac}

\affil[a]{\footnotesize Department of Computational Sciences, Institute for Particle and Nuclear Physics, HUN-REN Wigner Research Centre for Physics, 29-33 Konkoly-Thege Mikl\'os Street, H-1121 Budapest, Hungary}
\affil[b]{\footnotesize Department of Statistics, Institute of Data Analytics and Information Systems, Corvinus University of Budapest, 8 F\H{o}v\'am Square, H-1093 Budapest, Hungary}
\affil[c]{\footnotesize Department of Quantitative Methods, Faculty of Business and Economics, University of Pannonia, 10 Egyetem Street, H-8200 Veszpr\'em, Hungary}

\date{\footnotesize \today}

\begin{document}
\maketitle
\input{article_abstract}
\input{article_body}

\section*{Acknowledgments}
\input{article_acknowledgments}

\appendix
\input{causality_time}
\input{calculations_example}

\bibliographystyle{unsrt}
\bibliography{gd}

\end{document}

%% file: article_abstract.tex
\begin{abstract}
\noindent
Temporally evolving systems are typically modeled by dynamic equations. A key challenge in accurate modeling is understanding the causal relationships between subsystems, as well as identifying the presence and influence of unobserved hidden drivers on the observed dynamics. This paper presents a unified method capable of identifying fundamental causal relationships between pairs of systems, whether deterministic or stochastic. Notably, the method also uncovers hidden common causes beyond the observed variables. By analyzing the degrees of freedom in the system, our approach provides a more comprehensive understanding of both causal influence and hidden confounders. This unified framework is validated through theoretical models and simulations, demonstrating its robustness and potential for broader application.
\end{abstract}

%% file: article_body.tex
\sloppy

\section{Introduction}

Studying causal interactions between systems through observations of the evolution of their variables has recently become an active research area. Reichenbach's common cause principle, when adapted to stochastic dynamic systems (cf.\ \cite{reichenbach1956direction}), states that if two subsystems are not independent, then they share a common cause. The original principle is about events and also assumes that they are positively correlated. In that formulation, the common cause can be one of the subsystems itself, leading to the common way the principle is often cited.

Although identifying causal relationships without direct intervention is generally considered unattainable, several techniques have been developed to expose specific types of causal links by making assumptions about the underlying system or data. The most well-known and widely used method is Granger causality (GC) \cite{shojaie2022granger}, which is based on the Wiener-Granger predictive causality principle. Granger's approach and its generalizations (e.g., transfer entropy \cite{barnett2009granger}) can be applied to stochastic dynamic systems but not to deterministic ones. The GC method is widely used, though it cannot reveal, if there is, a hidden common cause; instead, it indicates one- or bidirectional causality.

Sugihara's Convergent Cross Mapping (CCM) method \cite{clark2015spatial} belongs to the family of topological causality methods. These methods share a foundation in Takens' embedding theorem \cite{takens2002reconstruction} and its various generalizations, including a particularly influential one proposed by Sauer \cite{sauer1994reconstruction}. This family, the family of topological causality, is designed to analyze the interplay between deterministic dynamic systems but cannot handle stochastic ones or detect common causes. In this family a new method, the dimensional causality (DC) method has been developed in \cite{benkHo2024bayesian} and further improved in \cite{benkHo2018}.  The DC method is able to reveal of the existence of a hidden common cause of deterministic dynamic systems. To the best of our knowledge, this was the first method of its kind (uploaded to arXiv.org in 2018) with a complete and statistically rigorous analysis. 
Other promising methods in this area include those by Hirata \cite{hirata2010identifying} and Krakovská \cite{paluvs2018causality}. In particular, Malinski \cite{Mal} removes almost all restrictions on latent processes, even allowing for contemporaneous causation, except for cyclic ones. Malinski extends the labeling of the edges of classical directed acyclic graphs (DAGs), but the completeness, as defined there, does not necessarily imply that all relations are unambiguous. Here, completeness means that all possible Markov equivalent maximal ancestral graphs (MAGs) can be generated, and no further structural information can be obtained from independence tests. Separately, the PCMCI method \cite{Runge2019detecting} is worth mentioning. One attempt to provide a full causal discovery of time series can be found in Mastakouri et al. \cite{mastakouri2021}, which assumes several strict conditions, but some of them are impossible to verify.

Further refinements and extensions have been developed in a series of papers by Runge and his coauthors, all based on conditional independence and Pearl's causal discovery algorithm (see \cite{Runge2019inferring}, \cite{Runge2018}, \cite{Runge2019detecting}, \cite{runge_causal_2023}, \cite{gerhardus_characterization_2023}, \cite{gerhardus_high-recall_nodate}, \cite{gerhardus_lpcmci_2021}, \cite{debeire2024bootstrap}, \cite{gunther2023causal}). Instead of attempting an exhaustive review of the vast literature, we highlight some recent and noteworthy surveys and offer some observations to position our own work. The capabilities and limitations of causal discovery algorithms have been thoroughly investigated in seminal works and reviews such as \cite{P1}, \cite{Spirtes}, \cite{Zh08}, \cite{Zh08a}, \cite{Mal}, and \cite{Lin}. See also \cite{Runge2018}, \cite{Runge2019detecting}, \cite{Runge2019inferring}, \cite{Glymour2019review}, \cite{Guo2020survey}, \cite{Yao2021Survey}, \cite{Yuan2022data-driven}, \cite{Kelly2022review}, \cite{Zanga2022Survey}, \cite{runge2023rev}, \cite{gong_causal_2023}, and \cite{gendron_survey_2023}. Emerging methodologies include the use of (deep) neural networks \cite{wang2024deep, berrevoets2024causal, ali2024causality}, representational learning \cite{doehnercausal2024}, and large language models \cite{wan2024bridging, jiaocausal, sun2024exploring, liu2024large}.

Broadly speaking, these approaches can be categorized by whether they allow for intervention or are purely observational, the type of system they analyze (e.g., deterministic or stochastic), the assumed properties of the time series (e.g., stationarity, noise characteristics), and the foundational principles they rely on, such as the Wiener-Granger predictive causality principle, topological (and dimensional) causality (often based on time-delay embedding), Pearl's causal discovery method, score-based techniques, and recurrence maps. Additionally, the methods differ in their ability to target various causal relationships, including identifying the existence and direction of direct causality, detecting indirect causal chains, recognizing causality amidst shared observational noise, and identifying hidden or latent common causes (or drivers).

These observations prompt the consideration of the number of degrees of freedom as a key concept in this investigation. The work by Iwasaki and Simon~\cite{iwasaki1994causality} can be regarded as a precursor to this study. Their paper defines causality between components of self-contained deterministic equilibrium and dynamic systems and introduces the notion of causal ordering. While their investigation focuses on well-defined and known mathematical equations, our work adopts the perspective of an observer of dynamic systems, aiming to reveal the causal relationships between components. Thus, we now turn to the definition and study of the degrees of freedom for different subsystems.

We propose a unified causality analysis of stochastic and deterministic systems named degrees of freedom causality method (\textit{df}-causality or \textit{df}-method in short). We consider a set of interacting, real-valued time-dependent variables and aim to determine the interplay between them. The observations are made in discrete time with uniform time steps. The method will be presented on two series of observed variables, called $x$ and $y$, but it will be clear that it is applicable to more general settings as well. Regardless of their stochastic or deterministic nature, we are able to determine the causal relationship between the systems from which the series originate\footnote{We assume that none of them is simply a function of the other (none of the subsystems are ``slaves'' of another). This can be detected, although in some cases it can be a nontrivial task (cf.\ \cite{zl}).}. In particular, we are able to detect whether they are independent, whether one drives the other, or whether there is a third, unobserved system that causes or drives both the observed ones (this system is usually referred to as the hidden common cause or common driver).

We remark that distinguishing different causal situation is constrained by observational precision and the stochastic nature of the systems. This includes the accuracy of the estimated distribution functions, which depends on the amount of observed data, and the precision of the conditions, which comes from either the bin size or the finite digit precision on a computer.

To avoid certain subtleties, we assume that in the case of driving, the cause precedes the consequence by one-time unit and there is no multi-lag between them. We also assume that the system and its subsystems are irreducible (there are no autonomous subsystems within them). Finally, we have two assumptions on the information injection from one subsystem to the other. First, the injected information is not a white noise process (or an i.i.d.\ sequence in the case of discrete time). Second, for continuous state space, the transfer function is generic, has no special symmetry (in Takens' sense), and is invertible (this will be specified precisely later).

The remainder of this paper is organized as follows. In Section~\ref{sec:basic_definitions}, we introduce basic definitions, followed by the presentation of the supporting mathematical framework in Section~\ref{sec:Causal_relations}. A comparison of various causality analysis methods is provided in Section~\ref{sec:comparison}. The proposed method is outlined in Section~\ref{sec:SimpleModel}, offering a practical approach for determining the degrees of freedom and demonstrating its application through examples. A real-world data example is presented and analyzed in Section~\ref{sec:chickenegg}. Finally, conclusions are drawn in Section~\ref{sec:conclusions}.

In the appendix we provide a short overview about the relation of the causality and time reversal symmetry in Section~\ref{sec:causality}, and give the details of the calculations in the main text.


\section{Basic Definitions}
\label{sec:basic_definitions}

In this paper, we focus on deterministic and stochastic dynamical systems on finite or infinite discrete or continuous spaces. The dynamic variable is denoted by $x(t) \in \mathbb{M}$, where $\mathbb{M = R}^d$ or $\mathbb{Z}^d$. A deterministic dynamical system in continuous time is defined by the differential equation:
\begin{equation}
    \label{def:PDE}
    f_{cont,i}(x^{(o_x)}, \xi ) = 0,
\end{equation}
where $x^{(o_x)}$ represents the tuple $(x_1^{o_1}, \dots, x_d^{o_d})$, with
\begin{equation}
    x_i^{o_i} = \left( x_i, \frac{d}{dt}x_i, \dots,\frac{d^{o_i}}{dt^{o_i}}x_i \right).
\end{equation}
The quantities $o_i \geq 1$ are the highest derivatives appearing in the differential equation for the $i$th component. The range of the function $f_{cont}$ is $\mathbb{M}$, and it describes the time evolution of the $i$th component. The stochastic components, denoted by $\xi$, represent white noise variables, and their actual values are independent of the entire past of $x$. In the deterministic case, the process $\xi$ is constant.

We assume that all measurements are taken at discrete times with uniform time steps $\tau$. Thus, from a continuous function $x(t)$, we observe the time series $X = \{ x_n \}$, where $x_n = x(t_0 + n\tau)$. We frequently use time-delay coordinates, and for shorthand notation, we define $X_{k}^{l}$ for $l > k$ as:
\begin{equation}
    X_{k}^{l} = \left( x_{l}, \dots, x_{k} \right).
\end{equation}

The differential equation \eqref{def:PDE} can be integrated to determine $x(t)$ from the previous $o_x$ values, considering the stochastic components as well. This leads to a recursion relation for the corresponding time series $X$:
\begin{equation}
    \label{eq:recx0}
    x_n = f\left( X_{n-o_x}^{n-1}, \xi_n \right).
\end{equation}
This recursive form is general and applies to both continuous and discrete time variables, with $\mathbb{M} = \mathbb{Z}^d$ or $\mathbb{R}^d$.

The time embedding can be flattened by introducing $\bar{x}_{na} = x_{n-a}$ for $a \in \{0,1,\dots, o_x-1\}$. We can then rewrite the recursion as:
\begin{align}
    &\bar{x}_{n0} = f( \bar{x}_{(n-1)0}, \bar{x}_{(n-1)1}, \dots, \bar{x}_{(n-1)(o_x-1)}, \xi_n), \nonumber\\
    &\bar{x}_{n1} = \bar{x}_{(n-1)0}, \nonumber\\
    &\vdots \nonumber\\
    &\bar{x}_{n(o_x-1)} = \bar{x}_{(n-1)(o_x-2)}.
\end{align}
This shows that there is a first-order rewriting of the recursion as:
\begin{equation}
    \label{eq:recx01st}
    \bar{x}_n = F\left( \bar{x}_{n-1}, \xi_n \right) \equiv F_n\left( \bar{x}_{n-1} \right).
\end{equation}
This equation can be solved via a stochastic function composition:
\begin{equation}
    \label{eq:recx}
    \bar{x}_n = F_n \circ F_{n-1} \circ \dots \circ F_1\left( \bar{x}_0 \right) \equiv F^{\circ n}(\bar{x}_0).
\end{equation}

\subsection{Degrees of Freedom}

We are now in a position to define the number of degrees of freedom:
\begin{definition}
    The \emph{number of degrees of freedom} \textit{df} for a given component $x_i$, denoted by $df_i$, is the number of initial conditions that must be specified to determine the complete time evolution of that component, given a stochastic history.
\end{definition}

As seen, \textit{df} is defined for individual components. If all components are interconnected, then, as shown in \eqref{eq:recx01st}, the number of values required to continue the recursion is equal to the sum of the order  number of  the components, giving $df = \sum_{i=1}^d o_i$ for all components. However, this is generally only an upper bound, as there may be subsystems of the original time series $X$ that are self-contained and might work as the driver (cause) for other components. In such cases, the subsystem can evolve independently of the remaining coordinates, requiring fewer initial conditions than the entire system. Therefore, for each component, we have:
\begin{equation}
    df_i \le \sum_{i=1}^d o_i.
\end{equation}
The causal relations between subsystems depend on this factor, as we will demonstrate.

The degrees of freedom correspond to the number of effective variables or initial conditions necessary to describe the system's evolution properly. For example, consider the first-order partial differential equation:
\begin{equation*}
\dot{x} = f(x),
\end{equation*}
which is a first-order system with $df_x = o_x = 1$. Clearly, the solution is well defined if one initial value $x_0$ is provided. In stochastic cases, the degrees of freedom are still linked to the number of initial states that must be specified to define the evolution of the distribution of the system correctly. For instance, consider a $k$th-order discrete-time Markov chain:

\begin{equation*}
x_n = f\left( x_{n-1}, x_{n-2}, \dots, x_{n-k}, \xi_n \right),
\end{equation*}
where $\{ \xi_n \}$ is an i.i.d. sequence, and $\xi_i$ is independent of the past of $x$, denoted by $\{ x_j \}_{j=0}^{i-1}$.

\subsection{Assumptions}

To avoid complications, we assume that in causal relationships, the cause precedes the effect by one unit of time. 
Multi-lag cases can be treated using longer time-delay embeddings, but the technique remains the same. For a detailed discussion of various lag scenarios, see \cite{zl}. For cases involving separate external drives for $\mathcal{X}$ and $\mathcal{Y}$, as well as an i.i.d. common driver, we refer to \cite{s3}.

The information transfer in recursions can be represented as a directed graph where vertices correspond to variables at each time step, and edges are time-invariant. This condition implies that the subject of our investigation is the so-called ``summary graph''.  We assume that the directed graph is acyclic (cf. \cite{Mal}). Additionally, we assume that the dynamical systems are irreducible (i.e., no autonomous subsystems) and converge to a stationary state.

Furthermore, we assume that the support of the stationary measure is a compact set $A \subset \mathbb{R}^{o_x}$, which also serves as the distribution for the initial values $\left( x_1, \dots, x_{o_x} \right)$ for the process $x_t$. With this randomized initialization of the system, we can treat $x_t$ as random (vector-valued) variables, even though the dynamics are deterministic.

For continuous state spaces, we assume that the transfer and observation functions are continuous, generic, without special symmetry (in Takens' sense), and time-independent. These assumptions align with those in Stark \cite{stark1999delay}, and without further mention, we follow the same assumptions throughout the paper. Specifically, we assume that the deterministic systems are non-degenerate, meaning none of the variables collapse into a smaller number during time evolution. Following Stark \cite{stark1999delay}, we also assume that the deterministic processes are reversible. If $\mathcal{X}$ drives $\mathcal{Y}$, as in
\begin{eqnarray}
     x_n = f(x_{n-1}), \\
     y_n = g(x_{n-1}, y_{n-1}),
\end{eqnarray}
then $g$ is invertible, meaning there exists a function $h$ such that
\begin{eqnarray}      
     x_n = h(y_{n-1}, y_n).
\end{eqnarray}
Of course, for stochastic processes, such invertibility holds only after fixing the noise instance.

Note that we assume the information transferred from $\mathcal{X}$ to $\mathcal{Y}$ is not an i.i.d. process. In our earlier paper, where we studied causal inference between Markov chains \cite{s3}, we discussed cases where i.i.d. injection is permissible and also considered the presence of hidden, non-common drivers. However, such extensions are not covered in this work. Finally, we assume that the driving function is not a simple mapping from $x$, meaning that $g$ cannot be reduced to $y_n = g(x_{n-1})$ (or, in the stochastic case, $y_n = g(x_{n-1}, \eta_n)$). This assumption ensures that the driving mechanism between the systems is non-trivial and cannot be simplified to a direct mapping of one variable onto the other.

\section{Causal Relations}
\label{sec:Causal_relations}

Here, we present the mathematical framework that supports the statements made in the previous section.

\subsection{Definitions}

\begin{definition}
We say that $\mathcal{X}$ drives $\mathcal{Y}$ if they satisfy the following recursion relations:
\begin{align}
    &x_n = F_n(X^{n-1}_{n-o_x}) \nonumber \\
    &y_n = G_n(X^{n-1}_{n-o_x}, Y^{n-1}_{n-o_y}).
\end{align}
\end{definition}

Let us note that $F_n = F_{\xi_n}$ and $G_n = G_{\eta_n}$ are stochastic mappings, with $\xi_n$ and $\eta_n$ being i.i.d. random variables. We also require that both $x$ and $y$ have non-trivial contributions to the evolution of $\mathcal{Y}$. Formally, this means that there are no functions $H_n$ such that $y_n = H_n(X^{n-1}_{n-o_x})$ or $y_n = H_n(Y^{n-1}_{n-o_y})$.

Using the solution of the recursion, we have:
\begin{eqnarray*}
X_n &=& F^{\circ n}\left( X_{-o_x+1}^{0} \right), \\
Y_n &=& G^{\circ n}\left( X_{-o_x+1}^{0}, Y_{-o_y+1}^{0} \right),
\end{eqnarray*}
where $\circ n$ denotes $n$-fold composition, cf. \eqref{eq:recx} for the random functions $f\left( X, \xi \right)$ and $g\left( X, Y, \eta \right)$. The distribution of $X_n$ is well defined if $o_x$ initial values are given ($X_{-o_x+1}^{0} \in \mathbb{M}^{o_x}$), and similarly for $Y_n$, we need $X_{-o_x+1}^{0} \in \mathbb{M}^{o_x}$ and $Y_{-o_y+1}^{0} \in \mathbb{M}^{o_y}$.

The same reasoning shows that the distribution of $Y_n$ and the joint system $\left( X_n, Y_n \right)$ is well defined if $(o_x - k, o_y + k)$ initial conditions are provided, where $0 \le k \le o_x$. This implies that $Y$ can be determined entirely from $o_x + o_y$ initial conditions. Since an appropriate number of initial conditions fully determines $Y$, we do not need to assume the existence of $X$ if it is not explicitly observed, except for mathematical convenience.

It is clear that the processes can be started at any later time (or stopped and restarted) by setting $X_m = A_m \in \mathbb{M}^{o_x - k}$ and $Y_m = B_m \in \mathbb{M}^{o_y + k}$. In that case:
\begin{equation*}
\left( X_n, Y_n \right) \text{ is independent of } \left( X_1^{m-1}, Y_1^{m-1} \right)
\end{equation*}
for all $n > m$. This means that the system has the Markov property.

The number of degrees of freedom can thus be defined as the number of conditions (at the present moment) necessary to separate the future state from the past (i.e., the Markov property). This observation leads to several useful consequences for detecting causal relations between systems.

With this framework, we can now define the causal scenario involving a common driver (or common cause).

\begin{definition}
The system $\mathcal{Z}$ is a common driver of $\mathcal{X}$ and $\mathcal{Y}$ if it drives both, and there is no direct interaction between $\mathcal{X}$ and $\mathcal{Y}$.
\end{definition}

Formally, this means:
\begin{align}
    &x_n = F_n(X^{n-1}_{n-o_x}, Z^{n-1}_{n-o_z}) \nonumber \\
    &y_n = G_n(Y^{n-1}_{n-o_y}, Z^{n-1}_{n-o_z}),
\end{align}
again with stochastic functions $F_n$ and $G_n$, where $Z$ represents the hidden process.

It should be noted that if $Z$ is an i.i.d. input, then $X$, $Y$, and the joint system $(X, Y)$ are Markov chains (cf. \cite{s3}). We exclude such pathological situations from our discussion.

Additionally, situations, where unobserved processes drive $X$ or $Y$ separately, are indistinguishable from the above scenario. Therefore, the most general causal relations involve the following possibilities: $X$ and $Y$ are independent (denoted $X \perp\!\!\!\perp Y$), $X$ drives $Y$ (denoted $X \to Y$), $Y$ drives $X$ ($Y \to X$), both systems drive each other ($X \leftrightarrow Y$), or there is a common cause driving both $X$ and $Y$.

As discussed above, these scenarios can be distinguished by determining the necessary number of conditions required to ensure the Markov property. This leads to Table \ref{tab:rel1}, where the last three columns represent the required length of the conditions.\footnote{The integers $k, l, p_.$ are non-negative, with $k+l=o_x+o_y$ and $0 \leq p_. \leq o_.$.}

\begin{table}[htbp]
    \centering
\begin{footnotesize}
\begin{tabular}{|>{\raggedright\arraybackslash}p{0.01\linewidth}|l|l|l|l|l|}
\hline
& Processes &  & $x$ & $y$  & $(x, y)$ \\ \hline
$1$ & $\mathcal{X} \perp\!\!\!\perp \mathcal{Y}$ & $\Leftrightarrow$ & $\left(o_x, 0\right)$ & $\left(0, o_y\right)$ & $\left(o_x, o_y\right)$ \\ 
\hline
$2$ & $\mathcal{X} \to \mathcal{Y}$ only & $\Leftrightarrow$ & $\left(o_x - p_x, p_x\right)$ & $\left(p_x, o_x + o_y - p_x\right)$ & $\left(p_x, o_x + o_y - p_x\right)$ \\ 
\hline
$3$ & $\mathcal{X} \leftarrow \mathcal{Y}$ only & $\Leftrightarrow$ & $\left(p_y, o_x + o_y - p_y\right)$ & $\left(p_y, o_y - p_y\right)$ & $\left(o_x + o_y - p_y, p_y\right)$ \\ 
\hline
$4$ & $\mathcal{X} \leftrightarrow \mathcal{Y}$ & $\Leftrightarrow$ & $\left(k, l\right)$ & $\left(k, l\right)$ & $\left(k, l\right)$ \\ 
\hline
$5$ & $\exists \mathcal{Z}$, $\mathcal{X} \centernot \leftrightarrows \mathcal{Y}$ & $\Leftrightarrow$ & $(o_x + o_z - p_{x,z}, p_{x,z})$ & $(p_{y,z}, o_y + o_z - p_{y,z})$ & $(o_x + p_z, o_y + o_z - p_z)$ \\ 
\hline
\end{tabular}
\end{footnotesize}
    \caption{The pairs $(n_1, n_2)$ in the ``x'', ``y'' and ``(x,y)'' columns represent the minimal lengths of the conditions needed to ensure the Markov property.}
    \label{tab:rel1}
\end{table}
A simple comparison of the values in different rows shows that all of these cases are mutually exclusive and cover all possibilities. Consequently, the implications from left to right can be reversed to obtain equivalences, enabling us to draw conclusions from observations.

Table \ref{tab:rel1} represents the ideal case where there is no observational noise. In realistic situations, we have a finite amount of data, and not all cases in Table \ref{tab:rel1} are equally suitable for numerical analysis. Table \ref{tab:rel2} below presents the least error-prone conditioning.

\begin{table}[htbp]
    \centering
\begin{footnotesize}
\begin{tabular}{|>{\raggedright\arraybackslash}p{0.01\linewidth}|l|l|l|l|l|}
\hline
& Processes & & $x$ & $y$  &(x,y)\\ \hline
$1$ & $\mathcal{X} \!\perp\!\!\!\perp \mathcal{Y}$
& $\Leftrightarrow $ & $\left( 
o_{x} ,0\right) $ & $\left( 0,o_{y}\right) $  &$\left( o_x
,o_{y}\right) $\\ 
\hline
$2$ & $\mathcal{X} \rightarrow \mathcal{Y}$ only& $\Leftrightarrow $ & $\left( o_{x} ,0\right) $& $\left(o_x ,o_y\right) $&$\left(o_x ,o_y\right) $\\ 
\hline
$3$ & $\mathcal{X}   \leftarrow \mathcal{Y}$  only& $\Leftrightarrow $ & $\left(o_x ,o_y\right) $& $\left(0 ,o_y\right) $&$\left(o_x ,o_y\right) $\\ \hline
$4$ & $\mathcal{X}  \leftrightarrow \mathcal{Y}$
& $\Leftrightarrow $ & $\left(o_x ,o_y\right) $& $\left(o_x ,o_y\right) $&$\left(o_x ,o_y\right) $\\ \hline
$5$ & $\exists \mathcal{Z}$,  $\mathcal{X}  \centernot \leftrightarrows \mathcal{Y}$& $\Leftrightarrow $ & $(o_x+o_z,0)$& $(0,o_y+o_z)$& $(o_x+o_z-r_z,o_y+r_z)$\\ \hline
\end{tabular}
\caption{Pairs represent the ``optimal'' lengths of the conditions needed to ensure the separation of $X_n$ and $Y_n$ from their respective pasts. Here, $r_z = 0$ or $r_z = o_z$.}
\label{tab:rel2}
\end{footnotesize}
\end{table}

The results from Table~\ref{tab:rel1} and Table~\ref{tab:rel2} can be summarized in Table \ref{tab:rel3}, which refers to the relationship between the orders of the observed systems, including the joint process. Let $o_J$ denote the order of the joint process $j_n = (x_n, y_n)$.

\begin{table}[htbp]
    \centering
\begin{footnotesize}
\begin{tabular}{|>{\raggedright\arraybackslash}p{0.01\linewidth}|l|l|l|}
\hline
$1$ &  $o_x, o_y < o_J = o_x + o_y$ &  $\Rightarrow$ &  $\mathcal{X} \perp\!\!\!\perp \mathcal{Y}$ \\ 
\hline
$2$ &  $o_x  < o_y = o_J < o_x + o_y$ & $\Rightarrow$ & $\mathcal{X} \to \mathcal{Y}$ only \\ 
\hline
$3$ & $o_y < o_x = o_J < o_x + o_y$ & $\Rightarrow$ & $\mathcal{X} \leftarrow \mathcal{Y}$ only \\ 
\hline
$4$ & $o_x = o_y = o_J < o_x + o_y$ & $\Rightarrow$ & $\mathcal{X} \leftrightarrow \mathcal{Y}$ \\ 
\hline
$5$ & $o_x, o_y < o_J < o_x + o_y$ & $\Rightarrow$ & $\exists \mathcal{Z}$, $\mathcal{X} \centernot \leftrightarrows \mathcal{Y}$ \\ 
\hline
\end{tabular}
    \caption{The implications can be deduced from Table \ref{tab:rel1}.}
    \label{tab:rel3}
\end{footnotesize}
\end{table}

Let us review some paradigmatic cases. If $\mathcal{X} \to \mathcal{Y}$ and there is a common driver $\mathcal{Z}$, then as seen in the DC method (cf. \cite{benkHo2018} or \cite{benkHo2018}), the driving from $\mathcal{X}$ to $\mathcal{Y}$ can be detected, but the existence of the common driver remains hidden. We believe that detecting such a hidden common driver is impossible without additional knowledge about the systems.

Now, consider a hidden, unobserved driver $\mathcal{U}$ of $\mathcal{X'}$: $\mathcal{U} \to \mathcal{X'}$ and $\mathcal{X} \to \mathcal{X'}$, $\mathcal{X} \to \mathcal{Y}$. If we observe $\mathcal{X'}$ and $\mathcal{Y}$, then, in general, our method (as well as the DC method) may suggest the existence of a hidden common driver (i.e., $\mathcal{X}$), even though $\mathcal{X}$ drives $\mathcal{Y}$. However, if $\mathcal{U}$ is an i.i.d. process, then $o_u = 0$, and we correctly conclude $\mathcal{X'} \to \mathcal{Y}$, while the DC method may not without specific detection or elimination of the noise influence.

\section{Comparison of Methods}
\label{sec:comparison}

Table \ref{tab:com} lists the possible causal relations between two systems and compares the application domains and conclusions provided by Wiener-Granger Causality (WGC), Convergent Cross Mapping (CCM), the Dimensional Causality (DC) method, and the method proposed in this paper (df).

\begin{table}[htbp]
\begin{center}    
\begin{tabular}{|l|l|l|l|l|}
\hline
& $
\begin{array}{c}
\text{WGC} 
\end{array} 
$ & $
\begin{array}{c}
\text{CCM} 
\end{array}
$ & $
\begin{array}{c}
\text{DC} 
\end{array}
$ & $
\begin{array}{c}
\text{df}  
\begin{array}{cc} 
\end{array} 
\end{array} 
$ \\
 & st & det & det & st, det \\\hline \hline
$\nexists z, x \rightarrow y$ & $\checkmark $ & $\checkmark $ & $\checkmark $ & $%
\begin{array}{cc}
\checkmark  & \checkmark 
\end{array}%
$ \\ \hline
$\nexists z, x \leftarrow y$ & $\checkmark $ & $\checkmark $ & $\checkmark $ & $%
\begin{array}{cc}
\checkmark  & \checkmark 
\end{array}%
$ \\ \hline
$\nexists z, x \longleftrightarrow y$ & $\checkmark $ & $\checkmark $ & $\checkmark $ & $%
\begin{array}{cc}
\checkmark  & \checkmark 
\end{array}%
$ \\ \hline
$\exists z$, $\mathcal{X} \centernot \leftrightarrows \mathcal{Y}$ & $\longleftrightarrow $ & heuristic & $\checkmark $ & $%
\begin{array}{cc}
\checkmark  & \checkmark 
\end{array}%
$ \\ \hline 
\end{tabular}
\caption{The valid or invalid detection of different models presented by WGC, CCM, DC, and df methods.}
\label{tab:com}
\end{center}
\end{table}

In Table \ref{tab:com}, the column headers represent the methods, where ``st'' stands for stochastic systems and ``det'' for deterministic systems. The Wiener-Granger Causality (WGC) indicates a bidirectional causal relation if there is a common cause or confounder.

The CCM method, which belongs to the family of topological causality approaches, is based on Takens' embedding theory (cf. Sugihara \cite{sugihara2012} and Harnack \cite{harnack2017}). When correlation is observed without inferred direct causation, it suggests the existence of a hidden common driver. However, it's essential to recognize that a shared driver doesn't always lead to a linear correlation between the two influenced systems. In \cite{sugihara2012}, no quantitative evaluation is provided. Suppose other measures of commonality between systems are applied, such as mutual information, challenges arise, for example, requiring a significant amount of data with an unknown sampling distribution. In contrast, \cite{harnack2017} provides a quantitative measure of causal strength, but the detection of the hidden common cause still relies on the correlation between the observed time series, and the conclusions are not fully quantified.

Hirata's work \cite{hirata2010} is based on recurrent maps and also presents a technique for detecting hidden common drivers. In this approach, a series of t-tests are used based on the overlap of recurrences. However, the detection of the hidden common driver remains on the ``cannot be rejected'' side of the series of applied t-tests, making it statistically inconclusive.

Lastly, Krakovsk\'a's work \cite{krakovska2019} intends to detect hidden common causes following the idea of \cite{benkHo2018}  but highlights that a proper statistical test to draw qualitative conclusions from dimension estimates is still lacking, leading to a high rate of false positives in detecting circular relationships instead of unidirectional couplings. The paper \cite{krakovska2019} uses correlation dimension, in contrast to the earlier unpublished work \cite{benkHo2018}, which employs the analogous and  topology-invariant intrinsic dimension and , offersing quantitative conclusions.

\section{Method: practical approach to the \textit{df}-causality method}
\label{sec:SimpleModel}

In the previous sections, we defined the concept of the number of degrees of freedom (\textit{df}) and demonstrated its utility as a powerful method for uncovering causal relationships.

This section provides a practical approach for determining \textit{df}, followed by a demonstration of its application through a semi-analytic example and a simple real-life data case.

\subsection{Determining the Number of \textit{df}}

The determination of the number of degrees of freedom requires fixing specific values in the past of a subsystem. After setting the \textit{df} values, the distribution of the subsystem's present values becomes independent of the number of applied constraints.

In practice, it is impossible to constrain a variable to an exact value because the probability of finding a data series with an exact value is zero. Therefore, constraints must be applied over a range, or more generally, all configurations are weighted by some distribution over past elements. The constraints we impose take the form
\begin{equation}
    C^{(n)}_k = \delta(x_{n-k} -\xi) P_k(\xi),
\end{equation}
where $P_k$ represents the distribution. If $P_k$ is sharp, $x_{n-k}$ is fixed to a specific value. If $P_k$ is a window function with width $\sigma_W$, $x_{n-k}$ is constrained to a certain range. In general, $P_k$ can represent any distribution with variance $\sigma_W$. We are interested in the distribution of $x_n$ after imposing these constraints:
\begin{equation}
    p(x_n\,|\, C^{(n)}_{k_1}, \dots, C^{(n)}_{k_a}) = {\cal N} \int p(x_n) \prod_a \delta(x_{n-k_a} -\xi_a) P_{k_a}(\xi_a)\, d\xi_1 \dots d\xi_a,
\end{equation}
where ${\cal N}$ is a normalization factor.

Beyond a certain number of constraints, the resulting distribution becomes independent of additional constraints. Although in principle we could apply all kinds of constraints, in practice, we typically use general ones, such as a zero-mean uniform window function with width $\sigma_W$ or a Gaussian distribution.

While it is useful to study the entire distribution, for practical purposes, we often compute only the variance $\sigma$. Though it is conceivable that the distribution might change without affecting its variance, it is unlikely that it remains unchanged under different numbers of constraints and varying window widths. Therefore, variance can serve as a reliable proxy for evaluating changes in the distribution.

To understand how the $p(x_n)$ distribution behaves under various constraints, we study the double-parameter dependent variance $\sigma(\sigma_W, n_{constr})$, where $\sigma_W$ represents the variance of the constraint distribution and $n_{constr}$ denotes the number of constraints. This function offers a straightforward approach to deduce the number of degrees of freedom. If insufficient constraints are applied, the limit $\sigma_W \to 0$ depends on the number of constraints. By observing the "saturation" of $\sigma$ as a function of $\sigma_W \to 0$, the number of degrees of freedom is revealed. The number of saturated lines corresponds to the \textit{df} of the component.

This method will be demonstrated in the following sections, using a semi-analytic example and a simple real-life data example.

\subsection{Semi-analytic Example: Linear System with Noise}

Since only linear systems can be solved analytically, we select a linear stochastic recursion for demonstration. Nevertheless, the general principles remain valid, as the method does not rely on the linearity of the equations.

We consider a stochastic recursion of the form
\begin{equation}
    \label{eq:linearEoM}
    x_n = Q x_{n-1} + S \xi_n,
\end{equation}
where $x_n \in \mathbb{R}^N$, and $\xi_n$ are i.i.d. normal random variables in $\mathbb{R}^N$. The matrices $Q$ and $S$ are of dimensions $\mathbb{R}^N \times \mathbb{R}^N$. We begin the evolution from an initial condition $x_0$, and the distribution of $x_n$ follows a Gaussian with mean $\bar{x}_n$ and correlation matrix $C_n$.

For the expected value $\bar{x}_n = \mathbb{E}[x_n]$, we have
\begin{equation}
    \bar{x}_n = Q \bar{x}_{n-1}.
\end{equation}
We assume that $\bar{x}_n = 0$ after a sufficiently long time, as this is not central to our analysis.

The evolution of the correlation matrix is given by
\begin{equation}
    C_n = \mathbb{E}[x_n \otimes x_n] = Q C_{n-1} Q^T + S S^T.
\end{equation}
The stationary solution satisfies
\begin{equation}
    C = Q C Q^T + S S^T.
\end{equation}
Its solution is
\begin{equation}
    C = \sum_{n=0}^\infty Q^n S S^T Q^{Tn}.
\end{equation}
This series converges for generic $S$ only if $|\lambda_i| < 1$ for all eigenvalues of $Q$. This condition also ensures that the process becomes independent of the initial conditions, i.e., $\bar{x}_n \to 0$ as $n \to \infty$.

\subsection{Conditional Probability Distributions}

In the proposed method, we compute empirical conditional probability distribution functions, where conditions are imposed on certain components. For linear systems, constraints do not alter the normal distribution, and the conditional distribution remains Gaussian with a modified covariance matrix and mean.

In stochastic systems, conditions typically involve both dynamic variables and noise terms. Let the dynamics be described by $x \in \mathbb{R}^N$ and the noise by $\xi \in \mathbb{R}^{N_{noise}}$, with $\xi$ being i.i.d. normal random variables. A generic linear condition can be written as
\begin{equation}
    R x - z - U \xi = 0,
\end{equation}
where $z \in \mathbb{R}^M$ specifies the condition's fixed value, $R$ is an $M \times N$ matrix, and $U$ is an $M \times N_{noise}$ matrix.

In practice, we cannot impose sharp constraints because real data cannot satisfy conditions with infinite precision. Thus, the condition is modeled as a random variable:
\begin{equation}
    \label{eq:linconstr}
    R x - z - U \xi = \eta,
\end{equation}
where $\eta$ is a normal random variable with zero mean and some correlation matrix $C_W$ (representing the window width).

To compute the resulting distribution after imposing these conditions, we perform Gaussian integrals, yielding the following result (details are in \ref{sec:Appendix_condprob}):
\begin{equation}
    p(x) \sim e^{-\frac{1}{2} (x - \bar{x})^T C_x^{-1} (x - \bar{x})},
\end{equation}
where
\begin{align}
\label{eq:barCandbarx}
    C_x^{-1} &= C^{-1} + R^T (C_\xi + C_W)^{-1} R, \nonumber \\
    \bar{x} &= C_x R^T (C_\xi + C_W)^{-1} z,
\end{align}
and $C_\xi = U U^T$.

When dealing with a stochastic process, where $C_\xi \neq 0$, it is beneficial to select a window function whose width matches the standard deviation of the stochastic noise. That is, $C_W$ and $C_\xi$ should be of the same order of magnitude. If $C_W \ll C_\xi$, then the sum will be dominated by $C_\xi$, and statistical power is lost. On the other hand, if $C_\xi \gg C_W$, accuracy decreases.

In real data analysis, a trade-off arises: imposing stricter conditions reduces the available data, limiting the precision of variance estimates. Therefore, it is generally unwise to use a window function that is too narrow, as this may hinder the reliable estimation of relevant variance parameters.

This analysis applies to the practically significant case where we impose constraints on the past elements of the series $x_{n-k}$ with some lag $k$. Although the formulas are complex, they are straightforward to compute (refer to \eqref{eq:barCandbarx} for the covariance matrix and \eqref{eq:barx_for_past} for the expected value). For further details, see Appendix \ref{sec:Appendix_past}.

\subsection{Example System}

To demonstrate our method, we use a system with six components, represented by the matrix $Q$:
\begin{equation}
    Q = \left(
    \begin{array}{cccccc}
        c_x & -s_x & 0 & 0 & g & 0 \\
        s_x & c_x & 0 & 0 & g & 0 \\
        0 & 0 & c_y & -s_y & g & 0 \\
        0 & 0 & s_y & c_y & g & 0 \\
        0 & 0 & 0 & 0 & c_z & -s_z \\
        0 & 0 & 0 & 0 & s_z & c_z \\
    \end{array}
    \right)
\end{equation}
where
\begin{equation}
    c_a = \alpha_a \cos(\varphi_a), \quad s_a = \alpha_a \sin(\varphi_a), \qquad a = x, y \;\mathrm{or}\; z.
\end{equation}
The noise correlation matrix is chosen to be proportional to the identity matrix, $S = \sigma \bm{1}$.

This system consists of three subsystems: the $x$, $y$, and $z$ subsystems, each having two internal degrees of freedom. The $z$ subsystem operates independently with a closed dynamic. The dynamics of $x$ and $y$ are influenced by $z$, but they do not interact directly with each other.

Using our earlier terminology, $z$ is an independent subsystem that drives both $x$ and $y$, and $x$ and $y$ share a common cause:
\begin{equation}
    z \to x, \quad z \to y, \quad \exists z \quad x \centernot \leftrightarrows y
\end{equation}

Regarding the number of degrees of freedom, to fully determine the time evolution in $z$ (given a stochastic history), it is sufficient to fix the two internal components of $z_{n-1}$, meaning the \textit{df} is 2 (one pair). For $x$, it is not enough to fix its internal components alone, as it interacts with the $z$ subsystem. To determine the time evolution of $x$, we need to fix both its internal components and the $z$ initial conditions. Alternatively, two pairs in the past of $x$ can be fixed, which indirectly provides all the required information. The $y$ subsystem behaves analogously to $x$. In summary:
\begin{equation}
    df_x = df_y = 4, \quad df_z = 2.
\end{equation}

However, if we consider the $(x, y)$ subsystem, we find that it has a total of 6 \textit{df} (the entire system). Thus, we conclude that $df_x + df_y > df_{xy} > df_x, df_y$, indicating the presence of a common cause.

We will reveal the same information from a purely data-based analysis.

\subsubsection{Numerical Setup}

In the numerical examples, we used two parameter sets. In the "deterministic" set, a small noise term was applied to simulate a deterministic system with minimal observational noise:
\begin{equation}
    \sigma = 5 \cdot 10^{-4}, \;
    \varphi_x = 0.5, \,
    \varphi_y = 0.3, \,
    \varphi_z = 0.4, \,
    \alpha_x = \alpha_y = \alpha_z = 1 - 10^{-6}, \,
    g = 0.3
\end{equation}
In the "stochastic" set, we used a moderate noise level:
\begin{equation}
    \sigma = 0.01, \;
    \varphi_x = 0.5, \,
    \varphi_y = 0.3, \,
    \varphi_z = 0.4, \,
    \alpha_x = \alpha_y = \alpha_z = 1 - 10^{-4}, \,
    g = 0.1.
\end{equation}
These parameters were selected to produce meaningful results.

In the numerical study, we varied two parameters: the number of constraints and the observation window width $\sigma_W$. The constraints involve fixing both internal components of a subsystem to zero. The number of constraints refers to how many internal component pairs from the $n-1$, $n-2$, ..., $n-k$ time steps were fixed to zero.

\subsubsection{Observing the \texorpdfstring{$z$}{\textit{z}} Subsystem}

After setting up the system, we can perform observations. The simplest case is observing one of the $x$, $y$, or $z$ subsystems.

\begin{figure}[htbp]
    \centering
    \includegraphics[width=5cm]{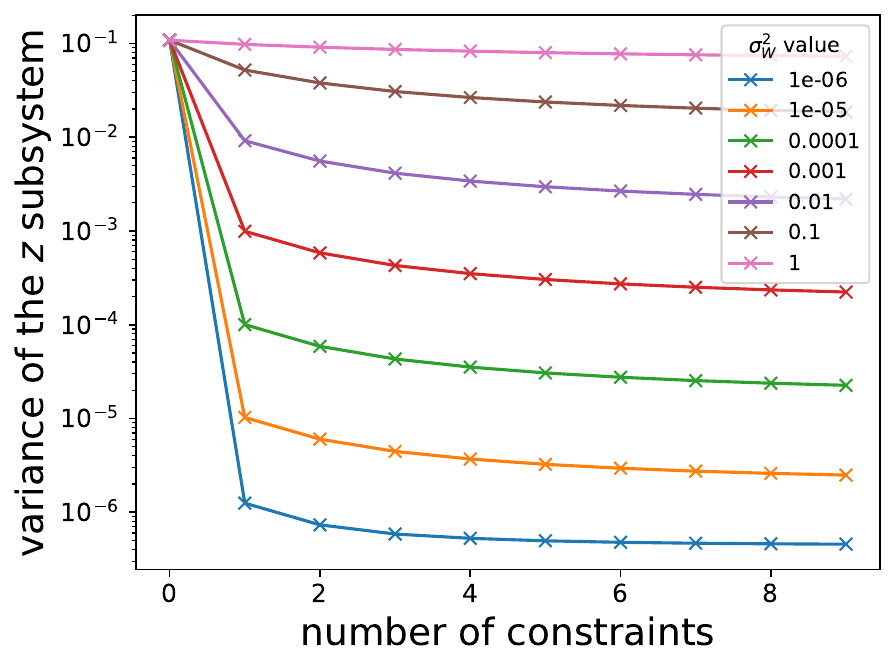}
    \includegraphics[width=5cm]{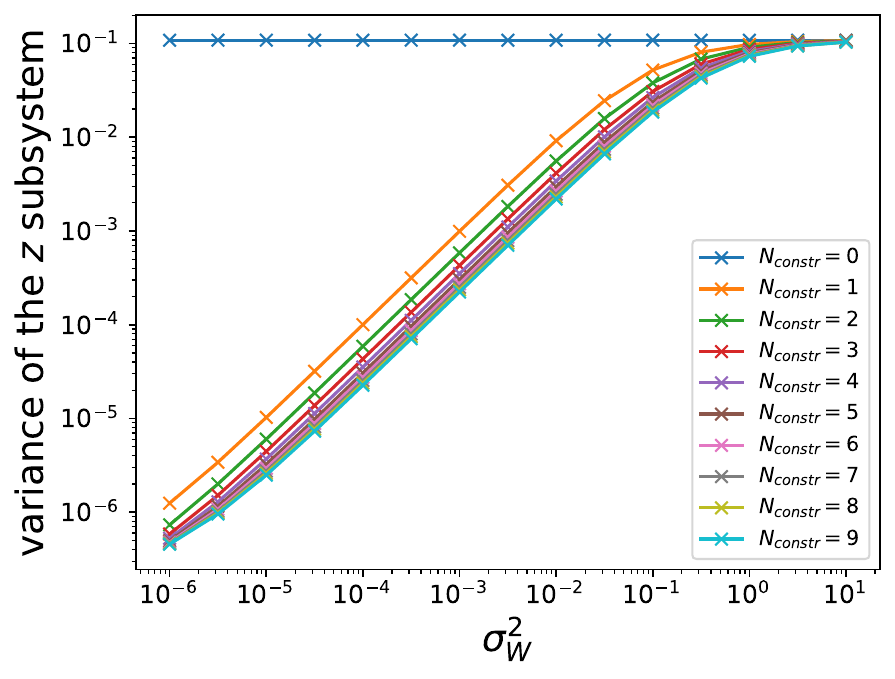}
    \includegraphics[width=5cm]{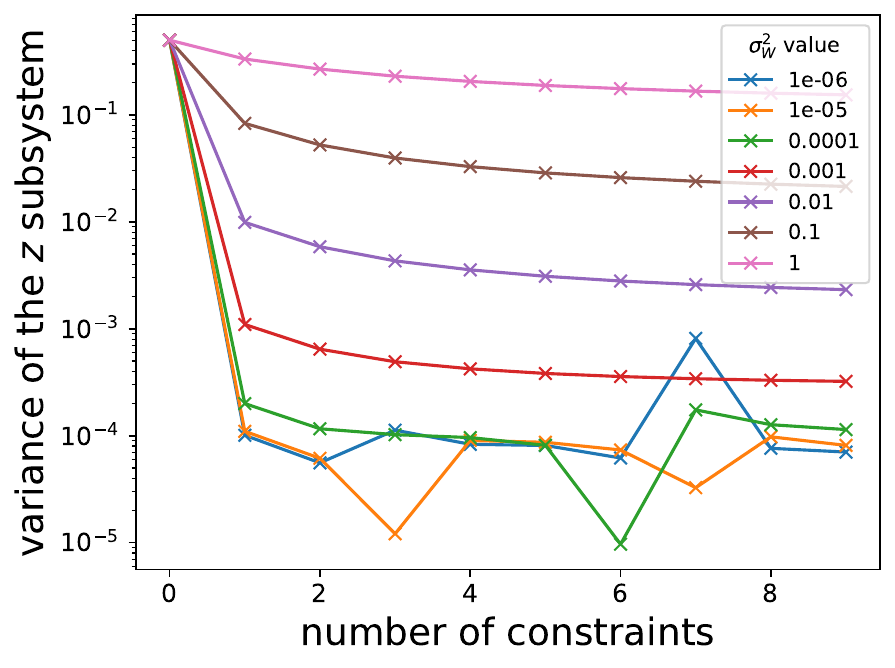}
    \includegraphics[width=5cm]{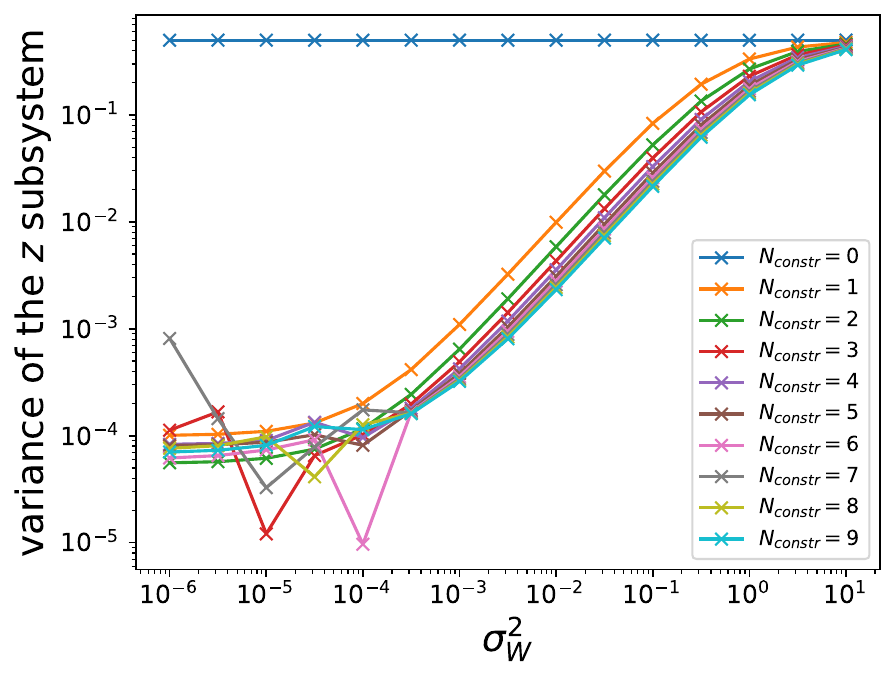}
    \caption{Observation of the $z$ subsystem: we apply different numbers of constraints using different bin widths (window widths). Top row: "deterministic" case; bottom row: "stochastic" case (for parameter assignments, see text).}
    \label{fig:c3v3}
\end{figure}

If we observe only the $z$ subsystem with different numbers of constraints and observation window widths, we obtain the plots in Fig.~\ref{fig:c3v3}.

In the top left panel, we observe how increasing the number of constraints on past values of $z$ narrows the distribution of the observed $z_n$ in the "deterministic" case. The first constraint has the most significant effect, but additional constraints continue to affect the $z_n$ distribution due to the finite width of the constraint window.

In the top right panel, the same relationship is shown, but this time plotted against the window width. As we can see, the variance of the observed distribution either remains constant (if no constraints are applied, $N_{constr}=0$) or scales proportionally to $\sigma_W$ in other cases.

This proportionality suggests that in the limit $\sigma_W \to 0$, the observed distribution approaches a Dirac delta function, meaning the value is fully determined. Indeed, by fixing the value of at least one pair in the past of $z$, the future values are fully determined.

This numerical analysis confirms that the \textit{df} of the $z$ subsystem is 2 (one pair): the minimal number of constraints needed to completely determine $z_n$. We can easily deduce the \textit{df} by counting the constant levels in the right panel.

The second row shows the same information in the presence of moderate stochastic noise. Here, the main difference is that below a certain window width (bin size), the observed distribution stabilizes: its width is determined by stochastic effects rather than the constraints. In practical applications, this is the optimal window width choice, as smaller widths do not provide further insights but reduce the available data.

We also note that increasing the number of constraints does not necessarily monotonically decrease the width of the observed distribution. Oscillations in the observed width can be seen in the plots, reflecting inherent characteristics of the system rather than numerical artifacts.

\subsubsection{Observing the \texorpdfstring{$x$}{\textit{x}} Subsystem}

\begin{figure}[htbp]
    \centering
    \includegraphics[width=5cm]{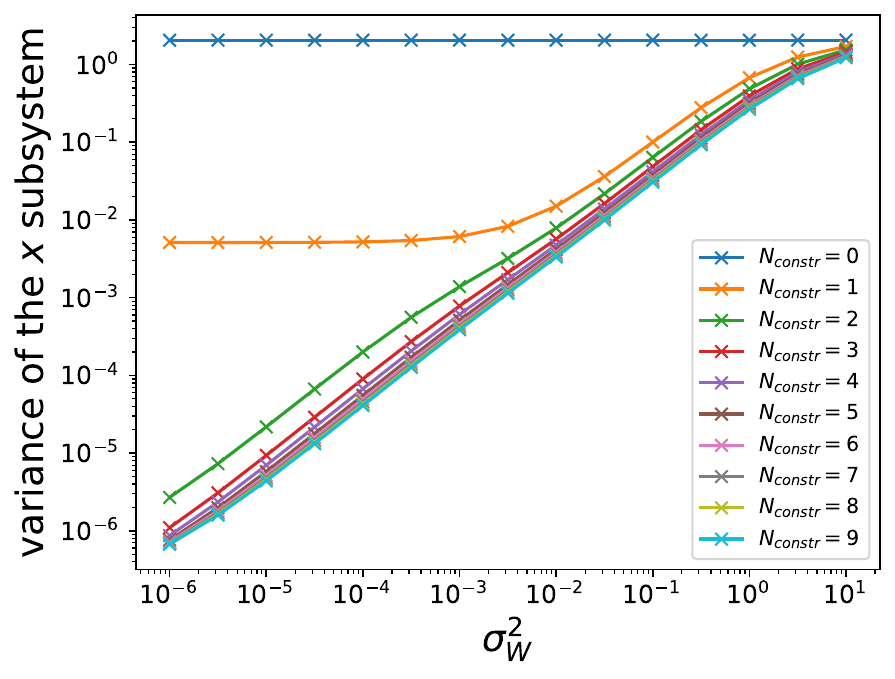}
    \includegraphics[width=5cm]{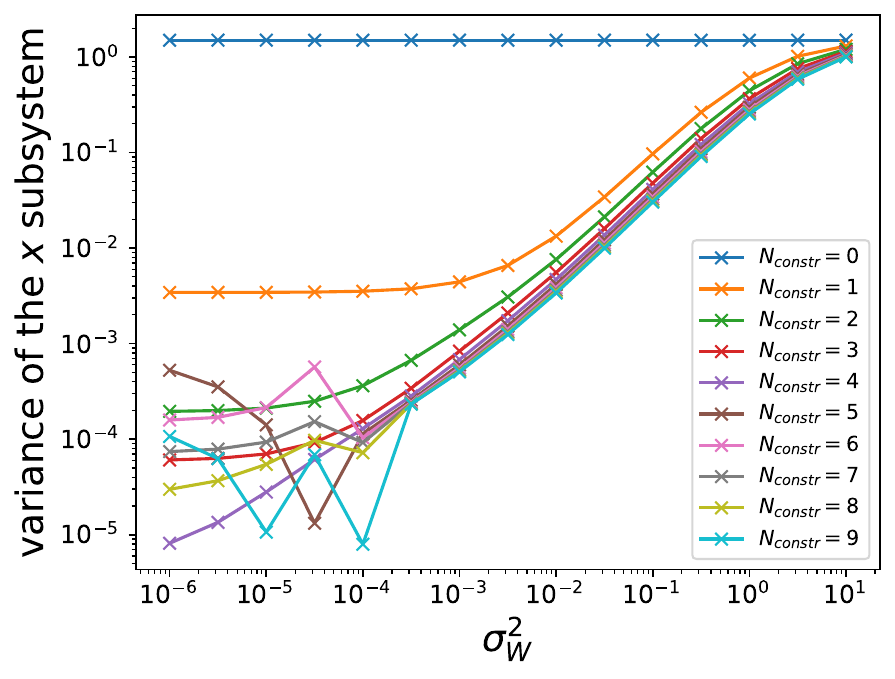}
    \caption{Observation of the $x$ subsystem: different numbers of constraints are applied with varying window widths. The left panel shows the "deterministic" case, and the right panel corresponds to the "stochastic" case (for parameter assignments, see text). The two stabilized lines indicate that this subsystem has 2 degrees of freedom (counting in pairs).}
    \label{fig:c1v1}
\end{figure}

We can perform the same observations with the $x$ subsystem. Below are the plots where the observed distribution width is plotted against the constraint window size (bin size), see Fig.~\ref{fig:c1v1}.

The primary difference between the $z$ and $x$ subsystems is that in the case of $x$, a single constraint on a past pair is insufficient to fully determine the observed value. Instead, at least two pairs of initial conditions are needed. This is clearly seen in the deterministic case (left panel of Fig.~\ref{fig:c1v1}), where two stabilized width lines do not converge to zero, indicating a \textit{df} of 4 (two pairs).

In the stochastic case (right panel), we can never observe a fully deterministic system (i.e., a Dirac delta distribution) due to the stochasticity. This is evident from the absence of lines that go linearly to zero as $\sigma_W \to 0$. Instead, all lines flatten below a certain constraint window width (bin size).

This plot highlights a phenomenon observed in all stochastic (i.e., real-world) systems: the influence from the $z$ subsystem can only be observed if there are three distinct flattened lines in the $\sigma$ versus $\sigma_W$ plot. Two of these lines correspond to the \textit{df}, while the third corresponds to the inherent stochasticity in the system. However, if the system is too noisy, two or more lines may merge, with the stochastic noise masking the external influence. In such a case, we may numerically conclude that the system is compatible with a \textit{df} of 2 (one pair), even though we know that there is an external influence.

This phenomenon is quite profound in real-world systems. For example, while we know that the position of Mars affects all systems on Earth due to its gravitational influence, the effect is too small to observe directly. It is overshadowed by numerous other effects of similar amplitude, which manifest as stochastic components in physical equations. Thus, only interactions significant enough to emerge from the stochastic background are considered relevant degrees of freedom.

\subsubsection{Observing the \texorpdfstring{$x$-$z$}{\textit{x}-\textit{z}} Subsystem}

When we observe both the $x$ and $z$ components, we can perform all the previous measurements, but we can also carry out new ones.

\begin{figure}[htbp]
    \centering
    \includegraphics[width=5cm]{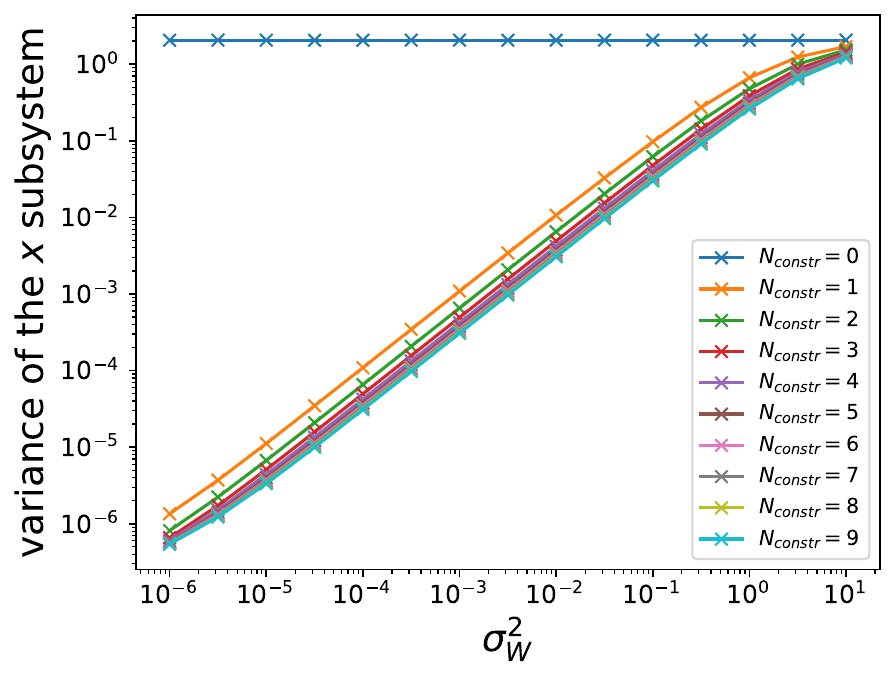}
    \includegraphics[width=5cm]{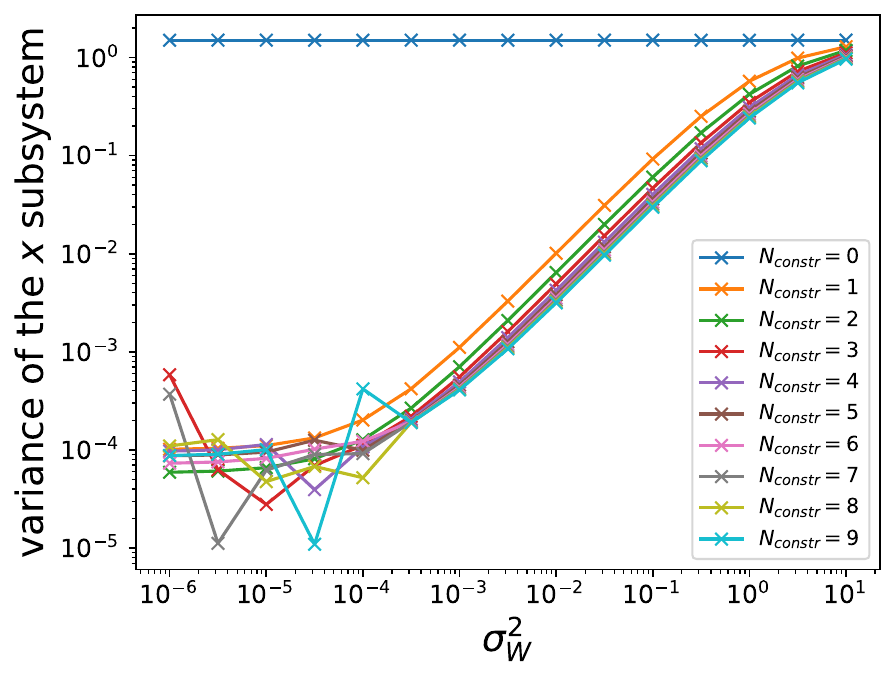}
    \caption{Observation of the $x$-$z$ subsystem: different numbers of constraints are applied with varying window widths, and we study the width of the observed $x$ distribution. The left panel shows the "deterministic" case, and the right panel shows the "stochastic" case (for parameter assignments, see text).}
    \label{fig:c13v1}
\end{figure}

The goal of these new measurements is to determine the extent to which the $x$-$z$ subsystem depends on external drivers. To do this, we can attempt to fix past values for both the $x$ and $z$ pairs. The corresponding figure can be seen in Fig.~\ref{fig:c13v1}.

These seemingly simple plots convey an important message: the $x$-$z$ subsystem is closed. By fixing a single value in the past, it is sufficient to fully determine the system (given a stochastic history). This confirms that the $x$-$z$ system is closed, with $df_x = 4$ and $df_z = 2$, which uniquely leads to the conclusion:
\begin{equation}
    z \to x.
\end{equation}

\subsubsection{Observing the \texorpdfstring{$x$-$y$}{\textit{x}-\textit{y}} Subsystem}

The most intriguing case arises when we observe the $x$-$y$ subsystem. In this scenario, we can replicate the earlier experiment, constraining the $x$ subsystem and analyzing the observed distribution width of the $x$ subsystem. A similar procedure can be applied to the $y$ subsystem, yielding analogous results. As a result, we determine that $df_x = df_y = 4$.

\begin{figure}[htbp]
    \centering
    \includegraphics[width=5cm]{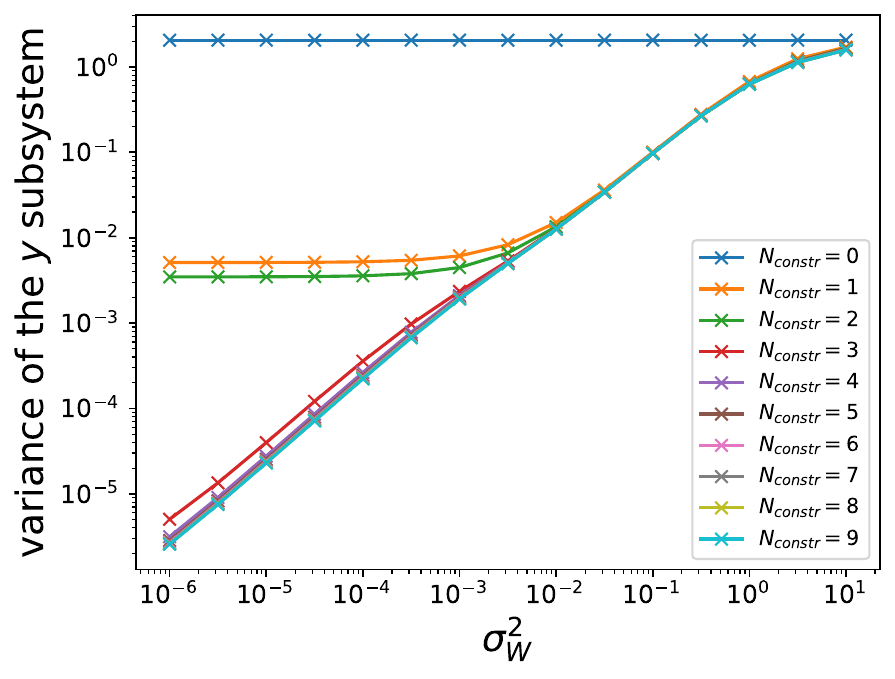}
    \includegraphics[width=5cm]{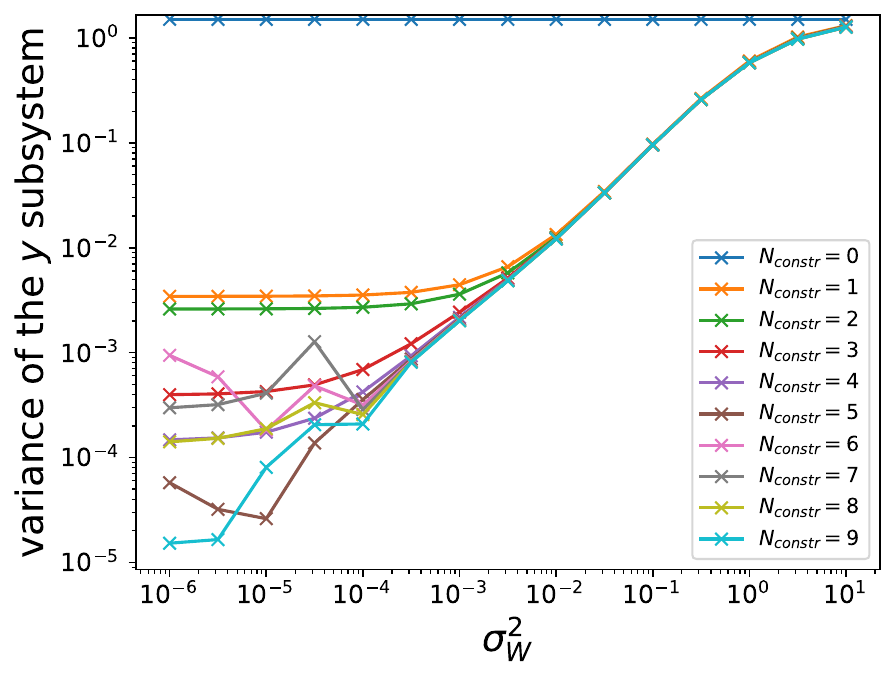}
    \caption{Observation of the $x$-$y$ subsystem: different numbers of constraints are applied using different bin widths (window widths), and we study the width of the observed $y$ distribution. The left panel corresponds to the "deterministic" case, and the right panel corresponds to the "stochastic" case (for parameter assignments, see text).}
    \label{fig:c12v2}
\end{figure}

The more interesting question concerns the \textit{df} of the combined $x$-$y$ subsystem. To investigate this, we adopted a specific constraint scheme:
\begin{itemize}
    \item For $N_{constr} = 0$, no constraints are applied.
    \item For $N_{constr} = 1$, the penultimate value of the $y$ subsystem (i.e., the $y_{n-1}$ pair) is constrained.
    \item For $N_{constr} > 1$, in addition to the above, $N_{constr}-1$ values of the $x$ subsystem (i.e., $x_{n-1}, \dots, x_{n-k}$ for $k=0,1,\dots, N_{constr}-1$) are constrained.
\end{itemize}
In this scheme, we examine the variance of the $y$ subsystem (we could also examine the $x$ subsystem with analogous results). The results are presented in Fig.~\ref{fig:c12v2}.

Both the deterministic and stochastic cases reveal that the $x$-$y$ system has a \textit{df} of 6 (three pairs). Recall that $df_x, df_y \leq df_{x,y} \leq df_x + df_y$, where equality on the left-hand side holds if at least one driving interaction occurs between the systems, and equality on the right-hand side holds if the systems are independent. If neither holds, then there is a common driver. In this case, the latter is true: we have a common driver, a common cause.

\section{Real-World Example: Chickens and Eggs}
\label{sec:chickenegg}

To demonstrate that our method works on fully experimental data, we analyzed the "chicken and egg" problem \cite{chegFisher}. The data describe egg production and the estimated chicken population in the US, based on U.S. Department of Agriculture data, for the period 1930–1980.

\begin{figure}[htbp]
    \centering
    \includegraphics[width=6cm]{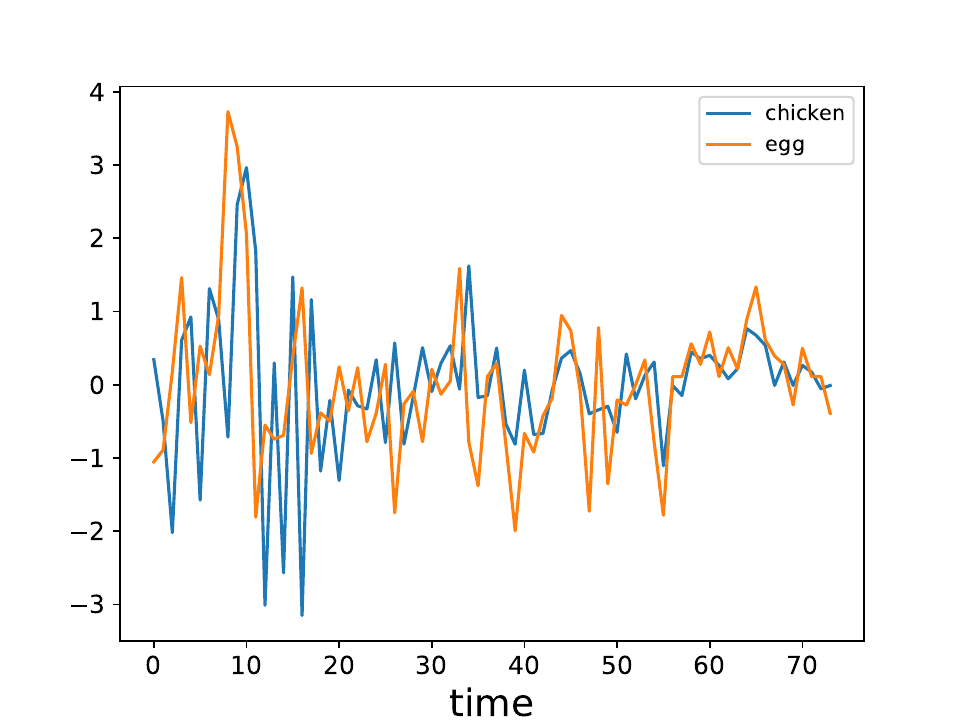}
    \caption{Chicken and egg data, normalized.}
    \label{fig:chicken-egg-data}
\end{figure}

To prepare the data for analysis, we differentiated both the chicken and egg datasets and normalized them to have unit variance. The resulting data is shown in Fig.~\ref{fig:chicken-egg-data}.

\begin{figure}[htbp]
    \centering
    \includegraphics[width=14cm]{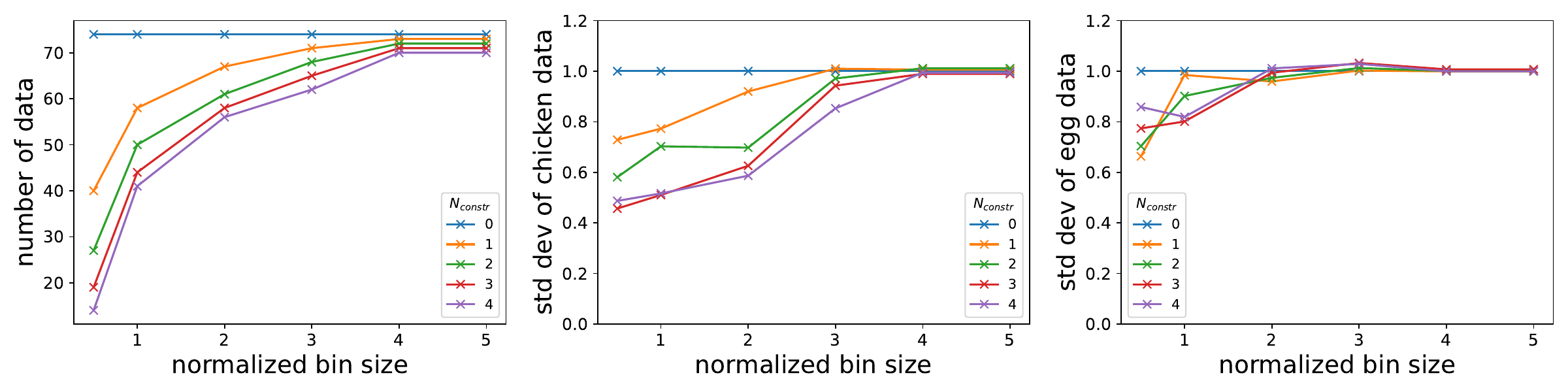}
    \caption{Results after constraining the chicken data. The left panel shows how much the amount of available data shrunk after applying the constraints. The middle and right panels plot the standard deviation of the chicken and egg distributions after imposing the constraint, respectively.}
    \label{fig:chicken_constrain}
\end{figure}

Next, we constrained either the chicken or the egg data using a finite-width bin and analyzed the width of the distribution of the same variables. The results of constraining the chicken data are shown in Fig.~\ref{fig:chicken_constrain}.

The middle panel illustrates that constraining the chicken data below a certain bin size affects its distribution, as expected. However, the right panel shows no observable change in the standard deviation of the egg data.

It is important to note that stricter constraints significantly reduce the amount of available data, as shown in the left panel of Fig.~\ref{fig:chicken_constrain}. This reduction implies that the standard deviation estimates become less reliable with decreasing bin size, which is reflected in the increased variance of the standard deviation estimates at small bin sizes (less than 1).

\begin{figure}[htbp]
    \centering
    \includegraphics[width=14cm]{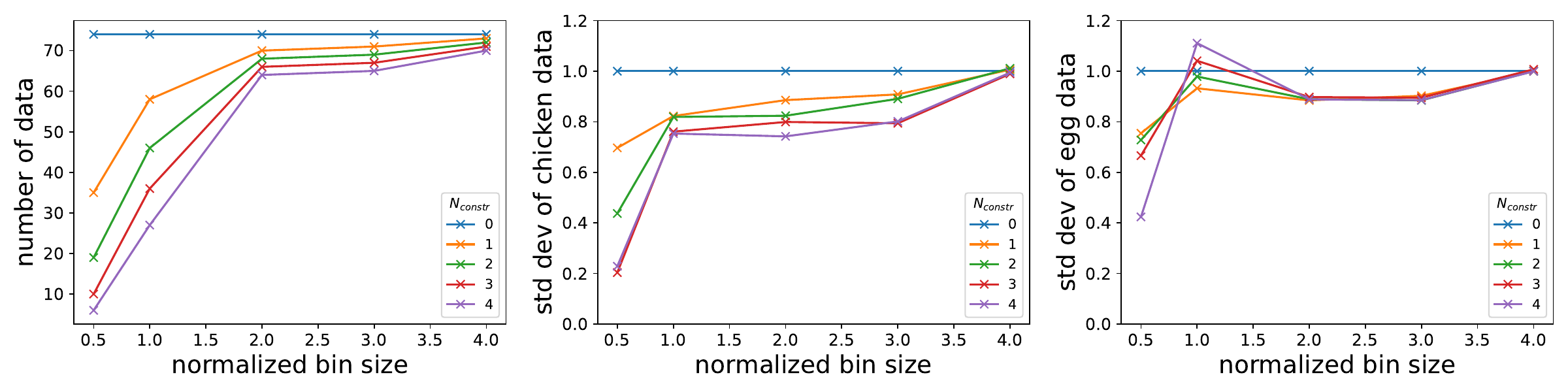}
    \caption{Results after constraining the egg data. The left panel shows how much the amount of available data shrunk after applying the constraints. The middle and right panels plot the standard deviation of the chicken and egg distributions after imposing the constraint, respectively.}
    \label{fig:egg_constrain}
\end{figure}

We can also analyze how the distribution changes after constraining the egg data. The results are presented in Fig.~\ref{fig:egg_constrain}.

The left and right panels reveal the expected results: the data amount is reduced as constraints are applied, and the egg distribution depends on the constraints imposed on the egg data.

However, the middle panel demonstrates an interesting result: the chicken data distribution also depends on the constraints applied to the egg data! This finding suggests that eggs drive the chicken population, at least based on these numerical results:
\begin{equation}
    \mbox{egg} \to \mbox{chicken},
\end{equation}
which aligns with the conclusions of the Granger causality analysis on this data \cite{chegFisher}. 

It is worth noting that a recent study \cite{chegBrazil} on chicken and egg production data from Brazil concluded a bidirectional relationship. However, the analysis of the joint system in this case is unnecessary, as common causality cannot be detected when direct causation between the systems is already present.

\section{Conclusions}
\label{sec:conclusions}
 
Several frameworks exist for time series causality analysis, including Granger causality, topological causality, and information causality, as well as variations and extensions like PCMCI and LPCMCI. These are among the most prominent methods (see \cite{granger1969investigating}, \cite{sugihara2012}, \cite{schreiber2000measuring}, \cite{runge2015identifying}). Most of these frameworks, however, focus exclusively on either deterministic or stochastic systems. Furthermore, only a few methods attempt to reveal the presence of hidden confounders, such as common drivers or common causes (see \cite{hirata2010}, \cite{krakovska2019}, \cite{gunther2023causal}, \cite{benkHo2024bayesian}). In this work, we have introduced a unified theoretical framework capable of addressing both challenges. At the core of this framework is the concept that investigating the degrees of freedom of a system, based on observed data, provides key insights. The degrees of freedom represent the minimum number of intermediate states required to make the past and future conditionally independent.

We also introduced an analytically tractable toy model. This model allows us to calculate the theoretical predictions of our method, simulating a scenario with an infinite time series, which provides optimal inference. To validate our theoretical framework and method, we simulated this toy model and calculated the degrees of freedom, while also performing robustness tests with simulated measurement errors. The results aligned perfectly with our theoretical expectations.

Finally, we applied our method to the millennia-old question: "Which came first, the chicken or the egg?" Our findings confirm that the egg came first, as suggested by previous work \cite{chegFisher}.

We believe that our method offers a valuable new tool for research in various fields. However, we recommend that the method be further tested on more synthetic and real-world datasets. In addition, we suggest that the current visual inspection of results be complemented by rigorous statistical analysis.

%% file: article_acknowledgments.tex
The research was supported by the Ministry of Innovation and Technology HUN-REN Office within the framework of the MI-LAB Artificial Intelligence National Laboratory Program. Project no. PD142593 was implemented with the support provided by the Ministry of Culture and Innovation of Hungary from the National Research, Development, and Innovation Fund, financed under the PD\_22 ``OTKA'' funding scheme.

%% file: causality_time.tex
\section{Causality and Time-Reversal Symmetry}
\label{sec:causality}

\noindent
The issue lies with time-reversal symmetry: in most microscopic physical systems, the dynamics are invariant under the reversal of the direction of time (time reflection symmetry\footnote{In a general local, causal, relativistic quantum field theory, the $CPT$ transformation is always a symmetry, where $C$ represents charge conjugation, $P$ spatial reflection, and $T$ time reflection.}). This implies that the concepts of ``earlier'' or ``later'' states are meaningless, and thus, the notion of ``causation'' becomes irrelevant. For example, if a change in variable $A$ causes a subsequent change in variable $B$, then by time-reversal symmetry, a change in $B$ would also cause a later change in $A$. As a result, their respective causal relationships cannot be defined. This sharply contrasts with causality in real life, where rain causes wet soil, but wet soil does not cause rain.

This situation closely resembles the problem of the arrow of time, which is also meaningless in a time-reversal invariant system. The resolution of the apparent contradiction between the time-reversal symmetric microscopic world and the macroscopic world, which exhibits a well-defined arrow of time, lies in the fact that in the macroscopic world, only collective phenomena are observed. These phenomena exhibit large values only in specific configurations with particular correlations between individual modes, and they dissipate as the system approaches equilibrium. This implies that individual collective coordinates decrease over time, even though the complete system remains time-reversal invariant. Naturally, by observing the entire system, it is possible to reverse this dissipation by gathering tiny pieces of information distributed across many variables. However, if measurements are made with only \emph{finite precision}, information is lost, and the original state cannot be restored.

Thus, the philosophical foundation of both the arrow of time and causality lies in our restricted measuring ability, which encompasses both the number of observed variables and the accuracy of the measurements.

There are two ways to account for inaccuracy in our predictions. The first approach is to acknowledge that the equations, and therefore the solutions, are not entirely accurate. This strategy is commonly used in various fields of physics, where ``exact'' time evolution equations are discussed, yet they never fully describe the system. In simpler systems, this results in measurements fluctuating around the predicted value (noisy measurements). However, in more complex, chaotic systems, the real and predicted trajectories deviate significantly, and the predictive power of these ``exact'' equations is lost, leading to only statistical predictions being possible.

Another way to incorporate inaccuracy is to treat the omitted variables statistically. Explicit ``noise'' terms can be introduced to represent these modes as collective phenomena, transforming the differential equations into stochastic differential equations.

At a given precision, a finite number of variables are retained in a dynamic system. The ``number of degrees of freedom'' (which will be precisely defined in the next section) describes the amount of information required to predict the system's time evolution at a given precision. At infinite precision, everything is interconnected, and the number of degrees of freedom becomes infinite. However, at finite precision, the number of degrees of freedom remains finite. Thus, this concept is intimately related to restricted accuracy and, consequently, to causality itself.

Heuristically, a small number of degrees of freedom in a quasi-closed subsystem implies that the subsystem is informed about only a small part of the larger system. This has significant implications. If subsystem $x$ drives subsystem $y$, then all the information required to understand $x$ is also necessary to understand $y$. Therefore, the number of degrees of freedom for $y$ must be larger. This also means that knowledge of $x$ alone cannot fully determine $y$, as more information is required. Conversely, if $y$ is completely known, all the information required for $x$ is available, and $x$ can be determined. If $x$ and $y$ share a common cause, then knowledge of $x$ provides only partial information about $y$, and vice versa. In this case, the $x$-$y$ compound system has fewer degrees of freedom than the sum of the individual degrees of freedom of $x$ and $y$.

It should again be emphasized that this is true only up to a given level of precision. In reality, at infinite precision, the system is time-reversal invariant, and the number of degrees of freedom becomes infinite. By decreasing the precision, a finite number of degrees of freedom is obtained, and causal relations emerge. Thus, the concept of causality is dependent on the level of accuracy. It is conceivable, for example, that at a given precision, $x$ may appear to drive $y$, but at higher precision, the reverse may be true. As an example, suppose $x$ is a signal preceded by a small-amplitude pre-tail. In specific systems with nonlinear amplification mechanisms, even the pre-tail may induce a signal in another system, $y$. At low precision, where the pre-tail is unobservable, it may seem that the bump in $y$ always precedes the bump in $x$, leading to the assumption that $y$ drives $x$. However, at higher precision, it turns out that $x$ drives $y$. This type of behavior can manifest in nonlinear optical systems, leading to phenomena such as superluminal light propagation~\cite{wang2000gain}.

%% file: calculations_example.tex
\section{Calculations in the Linear Example}


\subsection{Calculation of the Distribution of a Constrained System}
\label{sec:Appendix_condprob}

Here, we compute the distribution function of a Gaussian-distributed random variable with zero mean and a covariance matrix $C$, under a linear constraint 
\begin{equation}
    Rx - z - U\xi = \eta.
\end{equation}
Here $R$ is an $M \times N$ matrix, $U$ is an $M \times N_{noise}$ matrix and $\eta$ is a normal random variable with zero mean and some covariance matrix $C_W$ (referred to as 'W' for window), which simulates the role of finite bin size.

In the stationary case, the variables $x$ have a covariance matrix $C$, so the conditional distribution function is
\begin{equation}
    p(x) \sim \int d\xi \, d\eta \, e^{-\frac{1}{2} x^T C^{-1} x - \frac{1}{2} \xi^T \xi - \frac{1}{2} \eta^T C_W^{-1} \eta } \delta(Rx - z - U\xi - \eta).
\end{equation}
We can simplify this formula by introducing new variables. For convenience, we use the following transformation:
\begin{equation}
    \xi = U^T (UU^T)^{-1} \zeta + U_\perp^T \zeta_\perp.
\end{equation}
Here, $\zeta$ has $M$ components, $\zeta_\perp$ has $N_{noise}-M$ components, and we choose $U_\perp$ such that $UU_\perp^T = 0$ (i.e., $U_\perp$'s columns are orthogonal to those of $U$). If the rows of $U$ are independent, $UU^T$ is invertible, and if the rows of $U_\perp$ are linearly independent, the combined matrix $(U^T, U_\perp^T)$ is also invertible. This transformation provides a bijection $\xi \to (\zeta, \zeta_\perp)$. With the new coordinates, the expression becomes:
\begin{equation}
    p(x) \sim \int d\eta \, d\zeta \, d\zeta_\perp \, e^{-\frac{1}{2} x^T C^{-1} x - \frac{1}{2} \zeta^T (UU^T)^{-1} \zeta - \frac{1}{2} \zeta_\perp^T S_\perp S_\perp^T \zeta_\perp - \frac{1}{2} \eta^T C_W^{-1} \eta} \delta(Rx - z - \zeta - \eta).
\end{equation}
The integral over $\zeta_\perp$ can be performed without affecting the result, and we are left with:
\begin{equation}
    p(x) \sim \int d\eta \, d\zeta \, e^{-\frac{1}{2} x^T C^{-1} x - \frac{1}{2} \zeta^T (UU^T)^{-1} \zeta - \frac{1}{2} \eta^T C_W^{-1} \eta } \delta(Rx - z - \zeta - \eta). 
\end{equation}
This shows that the role of the window function and the stochastic noise are exactly the same.

After performing the Gaussian integrals, we obtain:
\begin{equation}
    p(x) \sim e^{-\frac{1}{2} (x - \bar{x})^T C_x^{-1} (x - \bar{x})},
\end{equation}
where
\begin{align}
\label{eq:barCandbarx1}
    &C_x^{-1} = C^{-1} + R^T (C_\xi + C_W)^{-1} R, \nonumber \\
    &\bar{x} = C_x R^T (C_\xi + C_W)^{-1} z,
\end{align}
and $C_\xi = UU^T$.

It is important to note that in the limit $C_\xi + C_W \to 0$, the values of $C_x$ and $\bar{x}$ remain finite if $R$ is not invertible. To see this, observe that the conditional distribution becomes:
\begin{equation}
    p(x) \sim e^{-\frac{1}{2} x^T C^{-1} x} \delta(Rx - z).
\end{equation}
Changing variables,
\begin{equation}
    x = R^T y + R^T_\perp y_\perp = \bar{R} \bar{y},
\end{equation}
where the constraint $y = (RR^T)^{-1} z$ holds, leads to:
\begin{equation}
    \bar{x} = R (RR^T)^{-1} z.
\end{equation}
The correlation matrix can be expressed as:
\begin{equation}
    \label{eq:constrainedC_det}
    C_x = \mathbb{E}[(x - \bar{x}) \otimes (x - \bar{x})] = \bar{R} \Pi^{(M)} \mathbb{E}[\bar{y} \otimes \bar{y}] \Pi^{(M)} \bar{R}^T,
\end{equation}
where $\Pi^{(M)}_{ij} = \delta_{ij} \Theta_{i > M}$ is a projector onto the non-fixed components of $y$. This gives:
\begin{equation}
    C_x = \Pi_{R\perp} C \Pi_{R\perp},
\end{equation}
where
\begin{equation}
    \Pi_{R\perp} = \bar{R} \Pi^{(M)} \bar{R}^{-1}.
\end{equation}
This is a projector, with $\Pi_{R\perp} R = 0$, meaning it projects onto the linear subspace orthogonal to the vectors spanned by $R$.

This matrix $C_x$ is not invertible. The eigenvectors corresponding to the zero eigenvalues are those that span the constraints.

\subsection{Fixing Past Elements}
\label{sec:Appendix_past}

When the constraints come from fixing the past elements of the $x_n$ series, we can derive a specific formula. The condition is:
\begin{equation}
    \label{eq:pastcond}
    x_{n-k,a} = z_{ka} + \eta_{ka},
\end{equation}
where we express $x_{n-k,a}$ in terms of $x_n$ and the noise terms using the equation of motion. 
This yields:
\begin{equation}
    z_{ka} + \eta_{ka} = (H^k x_n - H^k S \xi_n - H^{k-1} S \xi_{n-1} - \dots - H S \xi_{n-k+1})_a,
\end{equation}
which can be rewritten as:
\begin{equation}
    z_{ka} + \eta_{ka} = \sum_{i=1}^N (H^k)_{ai} x_{ni} - \zeta_{ka}.
\end{equation}
The matrix $R$ is:
\begin{equation}
    R_{ka,i} = (H^k)_{ai},
\end{equation}
and the noise term is:
\begin{equation}
    \zeta_{ka} = (H^k S \xi_n + H^{k-1} S \xi_{n-1} + \dots + H S \xi_{n-k+1})_a,
\end{equation}
with a correlation matrix:
\begin{equation}
    C_{\zeta, ka,\ell b} = \mathbb{E}[\zeta_{ka} \zeta_{\ell b}] = \sum_{m=0}^{\min(k,\ell)-1} (H^{k-m} SS^T H^{T,\ell-m})_{ab}.
\end{equation}
Assuming $SS^T = \sigma^2$, we obtain:
\begin{equation}
    \label{eq:zetaMatrix}
    C_{\zeta, ka,\ell b} = \sigma^2 \sum_{m=0}^{\min(k,\ell)-1} \sum_{i=1}^N (H^{k-m})_{ai} (H^{\ell-m})_{bi}.
\end{equation}

According to \eqref{eq:barCandbarx1}, the variance of the constrained system is:

\begin{equation}
    (C_x^{-1})_{ij} = (C^{-1})_{ij} + \sum_{ka,\ell b} (H^k)_{ai}(C_K^{-1})_{ka,\ell b} (H^\ell)_{bj},
\end{equation}
where
\begin{equation}
    \label{eq:CKmatrix}
    C_{K,ka,\ell b} = \sigma^2_{W,ka} \delta_{ka,\ell b} + \sigma^2 \sum_{m=0}^{\min(k,\ell)-1} \sum_{i=1}^N (H^{k-m})_{ai} (H^{\ell-m})_{bi}.
\end{equation}

We only need the specific elements of this matrix that correspond to the past conditions we are fixing. So, we select the pairs $\{k_1 a_1, k_2 a_2, \dots, k_M a_M\}$ where conditions are imposed. Then, we define:
\begin{equation}
    C_{K,\alpha\beta} = \sigma_{W\alpha}^2 \delta_{\alpha\beta} +  C_{\zeta,k_\alpha a_\alpha, k_\beta a_\beta}, \qquad R_{\alpha i} = H^{k_\alpha}_{a_\alpha i}.
\end{equation}
Finally, we compute the $C_x$ matrix by inverting the following expression:
\begin{equation}
    \label{eq:Cx_for_past}
    C_x^{-1} = C^{-1} + R^T C_K^{-1} R,
\end{equation}
and the expected value of $x$ comes from:
\begin{equation}
    \label{eq:barx_for_past}
    \bar{x} = C_x R^T C_K^{-1} z.
\end{equation}